# Tunable Emergent Heterostructures in a Prototypical Correlated Metal

D. M. Fobes,[1] S. Zhang,[2] S.-Z. Lin,[3] Pinaki Das,[1,*] N. J. Ghimire,[1,§] E. D. Bauer,[1] J. D. Thompson,[1] L.W. Harriger,[4] G. Ehlers,[5] A. Podlesnyak,[5] R.I. Bewley,[6] A. Sazonov,[7] V. Hutanu,[7] F. Ronning,[1] C. D. Batista,[1,2] and M. Janoschek[1,†]

[1]MPA-CMMS, Los Alamos National Laboratory, Los Alamos, New Mexico 87545, USA
[2]Department of Physics and Astronomy, The University of Tennessee, Knoxville, Tennessee 37996, USA
[3]T-4, Los Alamos National Laboratory, Los Alamos, New Mexico 87545, USA
[4]NIST Center for Neutron Research, National Institute of Standards and Technology, Gaithersburg, Maryland 20899, USA
[5]QCMD, Oak Ridge National Laboratory, Oak Ridge, Tennessee 37831, USA
[6]ISIS Facility, STFC Rutherford Appleton Laboratory, Harwell Science and Innovation Campus, Chilton, Didcot, Oxon OX11 0QX, United Kingdom
[7]Institute of Crystallography, RWTH Aachen University and Jülich Centre for Neutron Science (JCNS) at Heinz Maier-Leibnitz Zentrum (MLZ), Lichtenbergstr. 1, D-85747 Garching, Germany

*Current address: Division of Materials Sciences and Engineering, Ames Laboratory, U.S. DOE, Iowa State University, Ames, Iowa 50011, USA
§Current address: Argonne National Laboratory, Lemont, Illinois 60439, USA
†Corresponding author: mjanoschek@lanl.gov

**At the interface between two distinct materials desirable properties, such as superconductivity, can be greatly enhanced,[1] or entirely new functionalities may emerge.[2] Similar to in artificially engineered heterostructures, clean functional interfaces alternatively exist in electronically textured bulk materials. Electronic textures emerge spontaneously due to competing atomic-scale interactions,[3] the control of which, would enable a *top-down* approach for designing tunable *intrinsic* heterostructures. This is particularly attractive for correlated electron materials, where spontaneous heterostructures strongly affect the interplay between charge and spin degrees of freedom.[4] Here we report high-resolution neutron spectroscopy on the prototypical strongly-correlated metal $CeRhIn_5$, revealing competition between magnetic frustration and easy-axis anisotropy—a well-established mechanism for generating spontaneous superstructures.[5] Because the observed easy-axis anisotropy is field-induced and anomalously large, it can be controlled efficiently with small magnetic fields. The resulting field-controlled magnetic superstructure is closely tied to the formation of superconducting[6] and electronic nematic textures[7] in $CeRhIn_5$, suggesting that *in-situ* tunable heterostructures can be realized in correlated electron materials.**

The role of interfaces in enhancing or creating functionality is two-fold; interfaces exhibit reduced dimensionality, which is known to significantly influence electronic, magnetic and optical properties.[8] Furthermore, crossed response functions can arise from the interplay of two distinct order parameters at the interface, and lead to entirely new properties. This is successfully utilized in *bottom-up* approaches to device design. For example, semiconductor heterostructures can be grown with clean, atomically flat interfaces, the basis for applications in electronics and quantum optics.[9] Due to the intrinsic coupling between various order parameters, heterostructures grown from strongly correlated electron materials are a promising path towards new generations of devices, as highlighted by recent discoveries.[1,2,8] However, despite some impressive initial success, controlling these interfaces remains a significant challenge, precisely due to the underlying complexity.[8] Interestingly, this complexity is also what holds the key to a *top-down* approach for realizing high-quality interfaces. The complex ground states of strongly correlated electron materials arise from the competition between two or more atomic-scale interactions, often leading to superstructures, which we propose to exploit as *intrinsic heterostructures*.

We show that heavy electron metals, *i.e.* prototypical strongly correlated electron materials, are exceptional model systems to investigate intrinsic heterostructures. Here a frustrated Ruderman-Kittel-Kasuya-Yosida (RKKY) exchange interaction between localized *f*-electrons, which frequently favors spiral order, directly competes with a substantial easy-axis anisotropy enabled by the large spin-orbit interaction of lanthanide-based materials. The minimal model describing this competition is the Axial Next Nearest Neighbor Ising (ANNNI) Hamiltonian,[5] which shows that the conflict of frustration and anisotropy is universally resolved via the formation of modulated superstructures with applications in hard and soft matter.

As illustrated in Fig. 1a-c, in heavy electron metals the formation of a magnetic superstructure may also have important consequences for the electronic ground state. The presence of an additional Kondo interaction favors screening of *f*-electron magnetic moments by conduction electrons leading to heavy electronic quasiparticles with an enhanced electronic density of states (DOS). Due to this strong coupling between spin and charge, the underlying magnetic superstructure will likely induce a spatially modulated electronic texture (Fig 1b, c). Given that the period $\lambda$ of the magnetic superstructure is highly sensitive to external control parameters, our top-down approach offers the advantage that the electronic heterostructure can be tuned *in-situ*.

We demonstrate that a surprisingly small magnetic field of 2 T induces a substantial uniaxial magnetic anisotropy in the magnetically-frustrated heavy electron material $CeRhIn_5$, resulting in the formation of a field-tunable magnetic heterostructure. $CeRhIn_5$ is a tetragonal antiferromagnet (AFM), with Néel temperature $T_N = 3.8$ K at ambient pressure and zero magnetic field. Increasing pressure enhances the Kondo interaction via a growing overlap of neighboring Ce 4*f* orbitals, eventually leading to the complete suppression of the Ce magnetic moments at a magnetic quantum critical point (QCP) at $P_c = 2.25$ GPa around which a broad superconducting dome emerges (Fig. 1d).[10] Remarkably, in $CeRhIn_5$, part of the superconducting phase is textured (TSC in Fig. 1d).[6] In strikingly similar fashion, the AFM phase may also be suppressed by magnetic field ***H*** resulting in a QCP at $H_c = 50$ T, regardless of field direction.[11] Near this QCP, a new phase unstable towards the formation of an electronic nematic texture was recently discovered for $H > H^* = 28$ T (Fig. 1e). An arbitrarily small in-plane field component breaks the rotational symmetry of the electronic structure suggesting a surprisingly large nematic susceptibility.[7]

Interestingly, small in-plane fields also break the rotational symmetry of the AFM state, suggesting that electronic and magnetic textures are indeed related. Due to magnetic frustration arising from competing antiferromagnetic nearest- (NN) and next-nearest-neighbor (NNN) RKKY exchange along the *c*-axis,[12] the AFM order at low fields (AFM I in Fig. 2a,c) is an incommensurate spin spiral propagating along the *c*-axis with propagation vector ***k***$_I$ = (1/2 1/2 0.297), which conserves in-plane rotational symmetry.[13] However, for ***H***⊥*c*, a spin-flop transition occurs above the critical field $H_c^{III} = 2.1$ T (cf. Fig. 2a),[14] where the Ce moments align perpendicular to ***H*** forming a commensurate collinear square-wave phase, with propagation vector ***k***$_{III}$ = (1/2 1/2 1/4), suggesting a large magnetic-field-induced in-plane easy-axis anisotropy.

To elucidate the role of this field-induced easy-axis anisotropy, we investigate the magnetic interactions of CeRhIn$_5$ using neutron spectroscopy. This reveals that the magnetic interactions of CeRhIn$_5$ for in-plane fields are remarkably well described by the effective spin model Hamiltonian

$$\mathcal{H} = \sum_{ij}\left[J_{ij}\left((1-\delta)S_i^x S_j^x + (1+\delta)S_i^y S_j^y\right) + \Delta J_{ij} S_i^z S_j^z\right] - g_J \mu_B h \sum_j S_j^x, \quad (1)$$

which is related to the ANNNI model.[5] $S_i$ in Eq. (1) is a spin-1/2 operator representing the effective magnetic moment of the $\Gamma_7^2$ crystal field doublet. We note that the Hamiltonian in Eq. (1) is valid for $\boldsymbol{H}$ applied in the tetragonal basal plane, and adopt the convention that $\boldsymbol{H} \parallel (1\bar{1}0)$, so that the easy-axis anisotropy is along (110) (Fig. 2c). Our previous $H = 0$ study[12] revealed that the magnetic excitations are accurately described by $\mathcal{H}(\delta = 0, h = 0)$ with only three exchange constants $J_{ij}$: a NN exchange in the tetragonal basal plane, $J_0$, and two NN and NNN exchange interactions along $c$, $J_1$ and $J_2$, that, in combination with an easy-plane anisotropy $\Delta > 0$ in the basal plane, generate the spiral ground state (cf. Fig. 2c). Two additional ingredients are required to include field dependence: a conventional Zeeman term (final term in Eq. (1)) and a field-dependent easy-axis exchange anisotropy favoring spin alignment perpendicular to $\boldsymbol{H}$, described by the dimensionless parameter $\delta$. The above effective spin-1/2 Hamiltonian can be obtained by projecting the crystal field eigenstates onto the lowest energy doublet. The exchange anisotropy arises *a priori* from changes in the orbital character of the Ce 4$f$ electronic wave function with $H$, where its strength is expected to be substantial due to the large spin-orbit coupling for Ce and vary as $\delta(H) = I_\delta H^2$.

In Fig. 3, we show the full spin excitation spectrum of CeRhIn$_5$ as measured in the AFM III phase at $H = 7$ T (Fig. 2a, c), along the three principal directions ($h$00), ($hh$0), and (00$l$), centered at the commensurate magnetic zone center at $\boldsymbol{k}_{\text{III}} = (1/2\ 1/2\ 1/4)$, with additional fields presented in the Supplementary Information.[15] Comparing data sets at various magnetic fields reveals a clear field-induced increase in the spin gap $\Delta_S$ at $\boldsymbol{k}_{\text{III}}$. Fig. 4a presents $\Delta_S$ as function of $H$ extracted from energy cuts through the spin wave spectra shown in Fig. 3 at $\boldsymbol{k}_{\text{III}}$. The dynamic susceptibility $\chi''(\boldsymbol{q}, \omega)$ (cf. Fig 3d-f and Ref. [15]), and corresponding spin-wave dispersion is obtained from a large-$S$ expansion:

$$\hbar\omega_{\boldsymbol{q}} = \sqrt{\left(J_{0\boldsymbol{q}}^N + J_{2\boldsymbol{q}}^N \pm |J_{1\boldsymbol{q}}|\right)^2 + \left(J_{0\boldsymbol{q}}^A + J_{2\boldsymbol{q}}^A \pm |J_{1\boldsymbol{q}}|\right)^2}. \quad (2)$$

Here $J_{1\boldsymbol{q}}$ and $J_{0,2\boldsymbol{q}}^{A,N}$ are the Fourier transformation of the exchange parameters (Eq. S10-S14 in Ref.[15]), each consisting of the exchange integrals $J_0$, $J_1$, and $J_2$, easy-plane anisotropy $\Delta > 0$, and easy-axis anisotropy $\delta$, introduced in Eq. (1).

The dashed lines in Fig. 3 illustrate exemplary fits of $\chi''(\boldsymbol{q}, \omega)$ to our data, performed for every $H$ showing that the Hamiltonian in Eq. (1) describes our data quantitatively.[15] Due to the small size of the magnetic Brillouin zone along the $c$ direction, Umklapp scattering occurs at the zone boundary, resulting in additional spin wave branches, $\hbar\omega_{\boldsymbol{q}\pm\boldsymbol{k}_{\text{III}}}$.[12] The easy-plane anisotropy was fixed to $\Delta = 0.82$, as determined at $H = 0$,[12] and assumed to be field-independent; additional fit details are provided in the methods section. The resulting size of $\delta$ and exchange integrals as a function of $H$ are shown in Fig. 4b, c. Within AFM III, the parameters change smoothly with $H$; $J_0$, $J_1$ and $J_2$ decrease, in agreement with decreasing bandwidth of the spectrum, and $\delta$ increases in accordance with the growing spin gap. We note that the ratio of $J_2/J_1$ remains unchanged for all fields, indicating that the magnetic frustration is not affected by the applied magnetic field. Finally, as demonstrated by the red solid line in Fig. 4b, we find $\delta(H) = I_\delta H^2$ with $I_\delta = 0.0013(1)\ 1/\text{T}^2$. This implies that the experimental critical exchange anisotropy at $H_c^{III}$ is $\delta_c = 0.0057(5)$. By comparison, the critical exchange anisotropy calculated via mean-field modeling of the Hamiltonian in Eq. (1),[15] $\delta_c^{MF} = 0.0091$, agrees well with the experiment, which is remarkable considering that our model assumes *f*-electron localization in CeRhIn$_5$,[15] and that the mean field treatment neglects the effects of quantum fluctuations. Although the gap $\Delta_S = \hbar\omega_{k_{III}} = \sqrt{2\delta(2J_0 + J_2)[(2J_0 + J_2)(1 + \delta + \Delta) - J_1\Delta]}$ is the clearest indicator of increasing uniaxial anisotropy, it is also sensitive to the field-dependent exchange integrals *J*. By inserting interpolated values for the exchange integrals and $\delta$ we obtain the dashed line in Fig. 4a, demonstrating that our fits to the dynamic susceptibility quantitatively describe the observed spin gap for $H > H_c^{III}$. We note that an unexpected, small spin gap $\Delta_S \approx 0.25$ meV was observed at $H = 0$, but likely represents the longitudinal (or Higgs) mode that arises due to Kondo screening of the Ce magnetic moments, as explained in Ref.[15] (this scenario assumes that there is still a gapless transverse mode). Recent neutron diffraction measurements demonstrate that the Ce magnetic form factor is significantly different from free Ce$^{3+}$ with a magnetic moment that is reduced by 41% with respect to the expectation from the crystal field ground state, suggesting that the Kondo interaction in CeRhIn$_5$ is indeed substantial, in agreement with this scenario.[13]

As we show now, CeRhIn$_5$ exhibits an instability towards the formation of highly-tunable modulated magnetic superstructures. Using the exchange constants shown in Fig. 4c, and $\delta_c^{MF}$, we obtain the theoretical temperature vs. magnetic field phase diagram for CeRhIn$_5$ shown in Fig. 2b, based on our spin Hamiltonian and a mean-field calculation.[15] In addition to the remarkable agreement with the experimental phase diagram, it reveals a prominent feature of the ANNNI model, namely that the superstructure period is highly-tunable in proximity to $T_N$.[5] Notably, critical magnetic fluctuations immediately below $T_N$ compete with the uniaxial anisotropy, which causes a softening of the pinning of the magnetic moments along (110), ultimately leading to a magnetic structure with moments *primarily* along (110), but with small components parallel to **H**. This high-temperature phase (AFM II) is represented by an elliptical helix in which the size of the moments is modulated (cf. Fig. 2a, c).[14] Our model predicts a change of the magnetic propagation vector $k_{II}$=(1/2 1/2 *l*) as a function of both *H* and *T*. For the ANNNI model, the temperature dependence is given by $\Delta l(T) \propto -1/\ln(T - T_c^{III})$,[5] with $l = ¼$ at $T = T_c^{III}$ (critical temperature between AFM II and III), and slowly approaching the value dictated by NN and NNN exchange interactions along *c*, $l = 0.297$ for $T \rightarrow T_N$. In Fig. 2d we show that $l(T)$ at $H = 3.5$ T, as determined via high resolution neutron diffraction, indeed changes logarithmically, illustrating the ease with which the superstructure period $\lambda = 2\pi/k$ may be tuned.

The instability towards this highly-tunable magnetic heterostructure is apparent throughout the entire temperature-field-pressure phase diagram with significant impact on material properties. Transport measurements show that the AFM II phase continues to exist at pressures approaching the QCP.[16] Further, even for $H = 0$, the magnetic propagation vector changes from *$k_1$*=(1/2 1/2 0.326) to *$k_2$*=(1/2 1/2 0.391) near the phase boundary between textured and bulk superconducting states (indicated by the arrows in Fig. 1d).[17] Here the textured superconductivity is suggested to arise due to the coexistence of *$k_1$* and *$k_2$* magnetic domains, where the superconductivity only nucleates in *$k_2$*.[6] This may be explained via the mechanism shown in Fig. 1a-c, where *$k_1$* and *$k_2$* magnetic superstructures each induce distinct electronic textures, however with only one of them being compatible with the superconducting order parameter. This notably highlights that the tunable period of the magnetic heterostructure in CeRhIn$_5$ enables to control material properties.

Similarly, invoking the mechanism discussed in Fig. 1a-c for the field-induced nematic phase (Fig. 1e), an underlying modulated magnetic superstructure may generate two-dimensional (2D) electronic layers, where the direction of the local magnetic moments establishes a preferential direction that breaks rotational symmetry within the 2D layers with respect to the underlying lattice. For CeRhIn$_5$, the large field-induced magnetic anisotropy identified here can be accessed by a slight tilting of the magnetic field away from the *c* axis (inset of Fig. 1e) to align the magnetic moments, providing a natural explanation for the observed large nematic susceptibility.

Quantum oscillation measurements report a crossover from a small to a large Fermi surface volume near both QCPs (Fig. 1d, e), suggesting enhanced coupling between spin and charge degrees of freedoms due to the Kondo interaction in their vicinity.[18,19] This may explain why the magnetic superstructures that are omnipresent throughout the entire phase diagram predominantly influence material properties near the QCPs. Finally, the observed large uniaxial anisotropy arises due to changes of the orbital character of the Ce 4*f* electronic wave function with magnetic field. Remarkably, it has been demonstrated previously in the family of materials Ce*M*In$_5$ (*M* = Co, Rh Ir), to which CeRhIn$_5$ belongs, that the orbital character of the 4*f* wave functions can be also controlled via chemical substitution or pressure.[20] This not only affords an intrinsic mechanism for alternatively tuning the uniaxial anisotropy by pressure, but clarifies the striking similarity of the phase diagrams as function of *H* and *P* (cf. Fig. 1d,e).

In conclusion, via the quantitative application of an ANNNI-based effective spin model, a notable first for a heavy electron metal, we have identified a simple mechanism to create highly-tunable emergent magnetic heterostructures in CeRhIn$_5$ via competing interactions. Through coupling of spin and charge degrees of freedom mediated via the Kondo effect this mechanism concurrently generates electronic textures that significantly influence material properties. These textures are akin to emergent electronic heterostructures that exhibit clean interfaces and can be tuned with great ease employing using external tuning parameters such as magnetic field or pressure. Our work demonstrates that strongly correlated electron materials are a promising route for top-down approaches to producing tunable and emergent heterostructures. Notably, because frustrated exchange is common to *f*-electron materials, and field-induced uniaxial magnetic anisotropy has been reported in various heavy electron materials,[21,22] the mechanism identified here may apply universally for heavy electron materials. Furthermore, other classes of strongly correlated electron materials such as high-$T_c$ copper oxide, iron pnictide, and ruthenate superconductors all exhibit electronic textures near magnetic QCP,[23-25] many of which exhibit instabilities towards incommensurate modulated magnetism,[26-29] where orbital effects[30] and/or magnetic frustration[31] have similarly been proposed to be their origin, suggesting intrinsic functional heterostructures may be realized more broadly.

**Methods**

*Sample preparation*: Neutron scattering measurements were all performed on a mosaic (~2.2 g) of 14 CeRhIn$_5$ single crystals grown via the In self-flux method. To mitigate the effects of high neutron absorption by Rh and In, individual crystals were polished to a thickness of < 0.6 mm along the crystallographic *c*-axis and glued to a thin Al plate using a hydrogen-free adhesive (CYTOP). This sample mosaic is well-characterized and was used in our previous neutron spectroscopy study.[12]

*Neutron spectroscopy*: Time of flight neutron spectroscopy measurements shown in Fig. 3 and the supplement were performed on two direct geometry spectrometers: the Cold Neutron Chopper Spectrometer (CNCS)[32] at the Spallation Neutron Source (SNS), for applied magnetic fields of 5 T and below, with incident neutron energy $E_i$ = 3.315meV, and the LET Spectrometer[33] at the ISIS pulsed neutron and muon source for applied magnetic fields above 5 T, with $E_i$ = 3.3meV. Energy resolution in both cases was estimated to be ~0.08 meV. Inelastic slices with subtracted background were generated using Horace and fit to the theoretical dynamic susceptibility using a least-squares method implemented in NeutronPy (http://neutronpy.github.io/). Background scans were obtained on the CeRhIn$_5$ sample at $T$ = 20 K. Detailed inelastic neutron scattering measurements of the gap, shown in Fig. 4a, were performed on the Spin Polarized Inelastic Neutron Spectrometer (SPINS), a cold-neutron triple-axis spectrometer at the NIST Center for Neutron Research (NCNR), using a 7 T magnet with a $^3$He-dipper. Constant-$q$ scans were obtained with fixed $E_f$ = 3.0 meV, 40' collimation before the sample, a 60' radial collimator after the sample, and a horizontally-focused 11-blade PG(002) analyzer. Higher order neutrons were filtered using a cold Be-filter. Error bars of the gap values reflect the combined fitting error and energy resolution estimated by the quasielastic linewidth as measured on a standard vanadium sample. Error bars shown in Fig. 4b and 4c reflect the standard errors resulting from least-squares fitting. Diffraction data shown in Fig. 2d were also obtained on SPINS by performing scans along the (00$l$) direction with $E_i$ = $E_f$ = 3.315meV, 20'-S-10' collimation in triple-axis mode, a flat monochromator, and a flat 3-blade analyzer. Peak positions shown in Fig. 2d were obtained from fitting scans to a single Gaussian with constant background, and error bars represent the combination estimated error and momentum resolution calculated with the Cooper-Nathans method implemented in NeutronPy.

*Calculation of phase diagram*: To obtain the phase diagram, we treat spins in Eq. (1) as classical spins, and then numerically minimize the free energy of Eq. (1). We first perform numerical annealing using the Markov chain Monte Carlo method,[34] which minimizes the chance of trapping in a metastable state. Subsequently we use the relaxation method to determine the state with minimal free energy. Because the ordering wave vector is temperature-dependent, we continuously change the system size in the $c$-direction from 4 to 80, and keep the solution with the lowest free energy.

*Data Availability Statement (DAS):* The data that support the plots within this paper and other findings of this study are available from the corresponding author upon reasonable request. The neutron spectroscopy raw data from the experiment performed at LET are available at https://doi.org/10.5286/ISIS.E.82355430. Data from experiments carried out at SPINS are available at ftp://ftp.ncnr.nist.gov/pub/ncnrdata/ng5/201610/Fobes/CeRhIn5_22425/ and ftp://ftp.ncnr.nist.gov/pub/ncnrdata/ng5/201509/Fobes/CeRhIn5/.

**Supplementary Information** is available in the online version of the paper.

**Acknowledgements** We acknowledge useful discussions with Ryan Baumbach, Christian Pfleiderer, Markus Garst, Matthias Votja, Peter Böni, and Jon Lawrence. Work at Los Alamos National Laboratory (LANL) was performed under the auspices of the U. S. Department of Energy. LANL is operated by Los Alamos National Security for the National Nuclear Security Administration of DOE under contract DE-AC52-06NA25396. Research supported by the U.S. Department of Energy, Office of Basic Energy Sciences, Division of Materials Sciences and Engineering under the project 'Complex Electronic Materials' (Material synthesis and characterization) and the LANL Directed Research and Development program (neutron scattering, development of the spin wave model, mean-field computation and development of analysis software). Research conducted at Oak Ridge National Laboratory's (ORNL) Spallation Neutron Source was sponsored by the Scientific User Facilities Division, Office of Basic Energy Sciences, US Department of Energy. Experiments at the ISIS Pulsed Neutron and Muon Source were supported by a beam time allocation from the Science and Technology Facilities Council. We acknowledge the support of the National Institute of Standards and Technology, U.S. Department of Commerce, in providing the neutron research facilities used in this work.

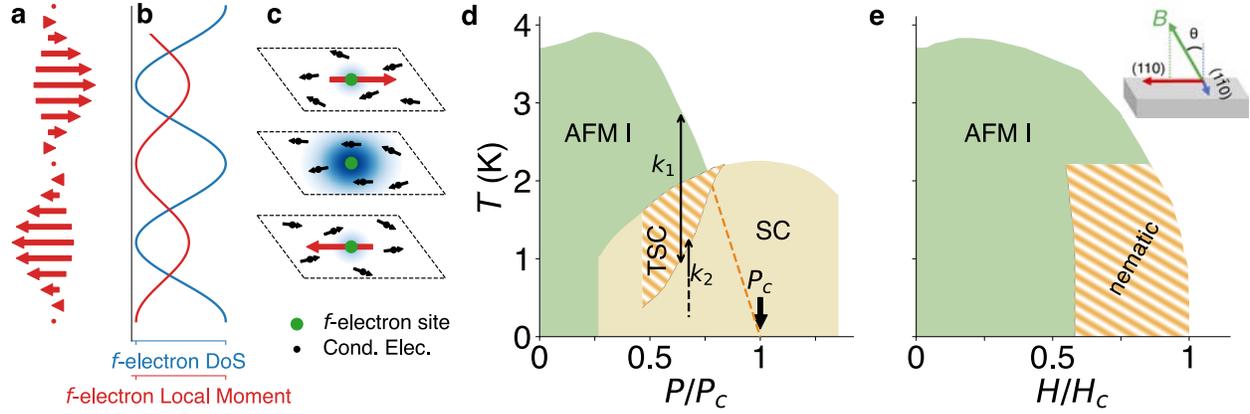

**Figure 1: Interplay of magnetic superstructures and electronic textures in heavy fermion materials. a** The competition of a frustrated Ruderman-Kittel-Kasuya-Yosida (RKKY) interaction, which typically promotes spiral order, is in direct conflict with a substantial easy-axis anisotropy enabled by the large spin-orbit interaction of lanthanide-based materials, and results in the formation of strongly modulated magnetic phases where the magnitude of the $f$-electron magnetic moment changes as a function of position. **b** Further, the Kondo interaction tends to hybridize conduction electrons and localized $f$-electrons by aligning conduction electrons spins antiparallel to $f$-moments. In the presence of strongly modulated $f$-electron moments this will generate an additional modulation of the $f$-electron contribution to the electronic density of states (DOS). **c** Illustration of extreme cases where the magnetic moment is maximum (top and bottom) and minimum (middle). The $f$-electron density of states at the Fermi level are represented by the blue-shaded region, where the electrons are more localized in the maximal moment case, and more itinerant in the minimal moment case. The prototypical heavy fermion material CeRhIn$_5$ investigated here exhibits two phases with electronic textures as shown in panel **d**, **e** that arise via the mechanism illustrated in **a-c** (see text). **d** Magnetic phase diagram as function of temperature $T$ and pressure $P$.[10] At ambient pressure and below Néel temperature $T_N$ = 3.8 K CeRhIn$_5$ orders antiferromagnetically (AFM I). Application of pressure suppresses the AFM I order resulting in a quantum critical point (QCP) at $P_c$ = 2.25 GPa around which a broad dome of unconventional superconductivity (SC) emerges. TSC denotes a region of textured superconductivity.[6] Arrows indicate temperature regions where magnetic ordering wave vector is $k_1$=(1/2 1/2 0.326) and $k_2$=(1/2 1/2 0.391), at $P \sim 1.48$ GPa.[17] **e** Magnetic phase diagram as a function of temperature $T$ and magnetic field $H$. The AFM I state can alternatively be suppressed at a second QCP by applying a critical field $H_c$ = 50 T. Magnetic fields applied with a small in-plane component results in the formation of an electronic nematic phase above $H^*$=28 T ($H^*$ varies slightly as a function of $\theta$, see Fig. 3 of Ref.[11]) for temperatures below $T$=2.2 K.

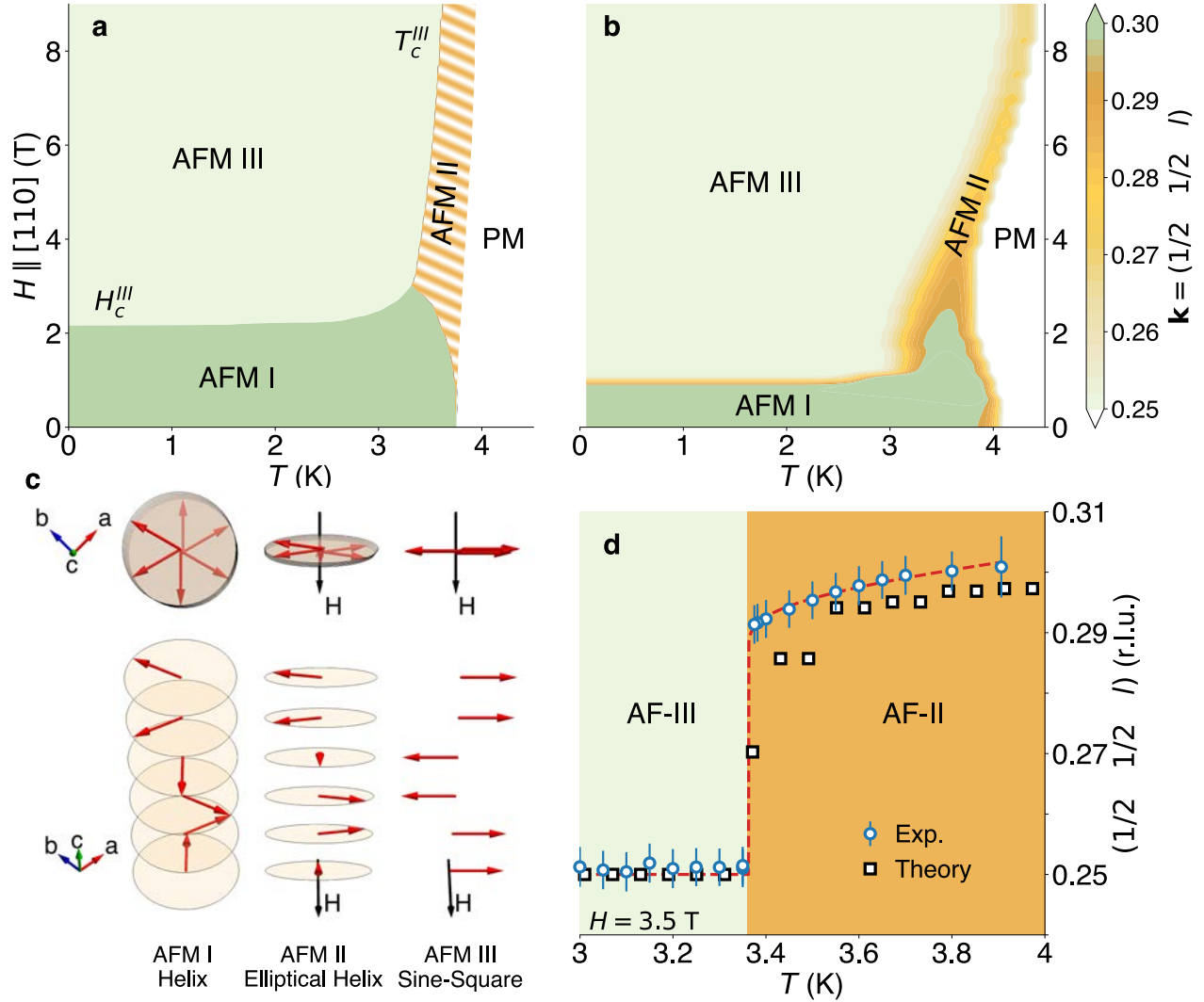

**Figure 2: Signatures of highly-tunable modulated magnetic superstructures in CeRhIn$_5$. a** Below $T_N = 3.8$ K at ambient pressure, CeRhIn$_5$ orders in an incommensurate antiferromagnetic spin helix (AFM I), with a propagation vector $k_I = (1/2\ 1/2\ 0.297)$, where the magnetic moments lie parallel to tetragonal basal plane.[13] Note that the AFM I phase conserves the four-fold rotational symmetry of the underlying crystal structure (see also **c**). Applying $H$ parallel to the tetragonal basal plane of CeRhIn$_5$ breaks the four-fold symmetry and results in the emergence of two additional magnetic phases: at high temperature, an incommensurate elliptical helix (AFM II) with strongly modulated magnetic moments and temperature-dependent propagation vector $k_{II} = (1/2\ 1/2\ l(T))$ (see also **d**) and at low temperature, a commensurate collinear square-wave (AFM III, "up-up-down-down" configuration) with a propagation vector $k_{III} = (1/2\ 1/2\ 1/4)$, separated from AFM I by critical magnetic field $H_c^{III}$, and from AFM II by critical temperature $T_c^{III}$.[14] **b** $T$ vs. $H$ phase diagram for CeRhIn$_5$ calculated based on our effective spin Hamiltonian, using the exchange interaction and field-dependent uniaxial magnetic anisotropy determined via neutron scattering. Color scale denotes the $c$-component of the magnetic propagation vector $k = (1/2\ 1/2\ l)$, derived from Eq. (1). **c** Illustrations of the three magnetic structures. Upper panels contain the projection of the three unit cells onto the tetragonal basal plane, clarifying the orientation of the

Ce magnetic moments (red arrows) in the plane. When magnetic field is applied in the tetragonal basal plane (here $H \parallel (1\bar{1}0)$, see black arrows), all (AFM III), or most (AFM II) Ce magnetic moments align perpendicular to $H$. Note that for the AFM II phase, the size of the Ce magnetic moment is strongly modulated. **d** The *c*-component of the magnetic propagation vector $k = (1/2\ 1/2\ l)$ at $H = 3.5$ T as a function of temperature from experiment and as calculated from Eq. (1), seen in the theoretical phase diagram in Fig. 1d. Dashed line indicates fit to logarithmic function $-1/\ln(T - T_c^{III})$. The logarithmic temperature-dependence of the propagation vector is characteristic of modulated superstructures as described Axial-Next-Nearest-Neighbor (ANNNI) framework,[5] and illustrates the highly tunable superstructure period $\lambda = 2\pi/k$. Note that for CeRhIn$_5$, $\lambda$ may be tuned as a function of $T$ or $H$ (see **b**). Error bars represent standard deviations.

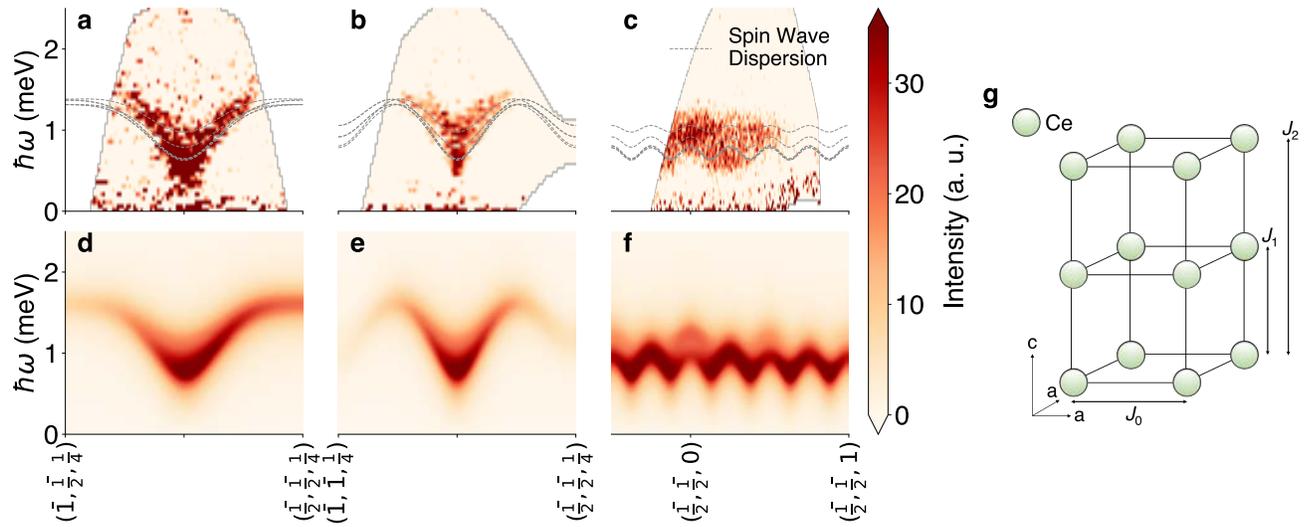

**Figure 3: Magnetic excitations of CeRhIn$_5$ in in-plane magnetic fields. a-c** Measured spin excitation spectra at $H = 7$ T in the AFM III phase where $k_{III} = (1/2\ 1/2\ 1/4)$, along three high symmetry directions: **a** ($h00$), **b** ($hh0$), **c** ($00l$). Dashed lines indicate spin wave dispersions (cf. Eq. (2)), resulting from fitting. **d-f** Calculated dynamic magnetic susceptibility $\chi''(q,\omega)$ using the fitted parameters. **g** Three magnetic exchange integrals used for our calculations.

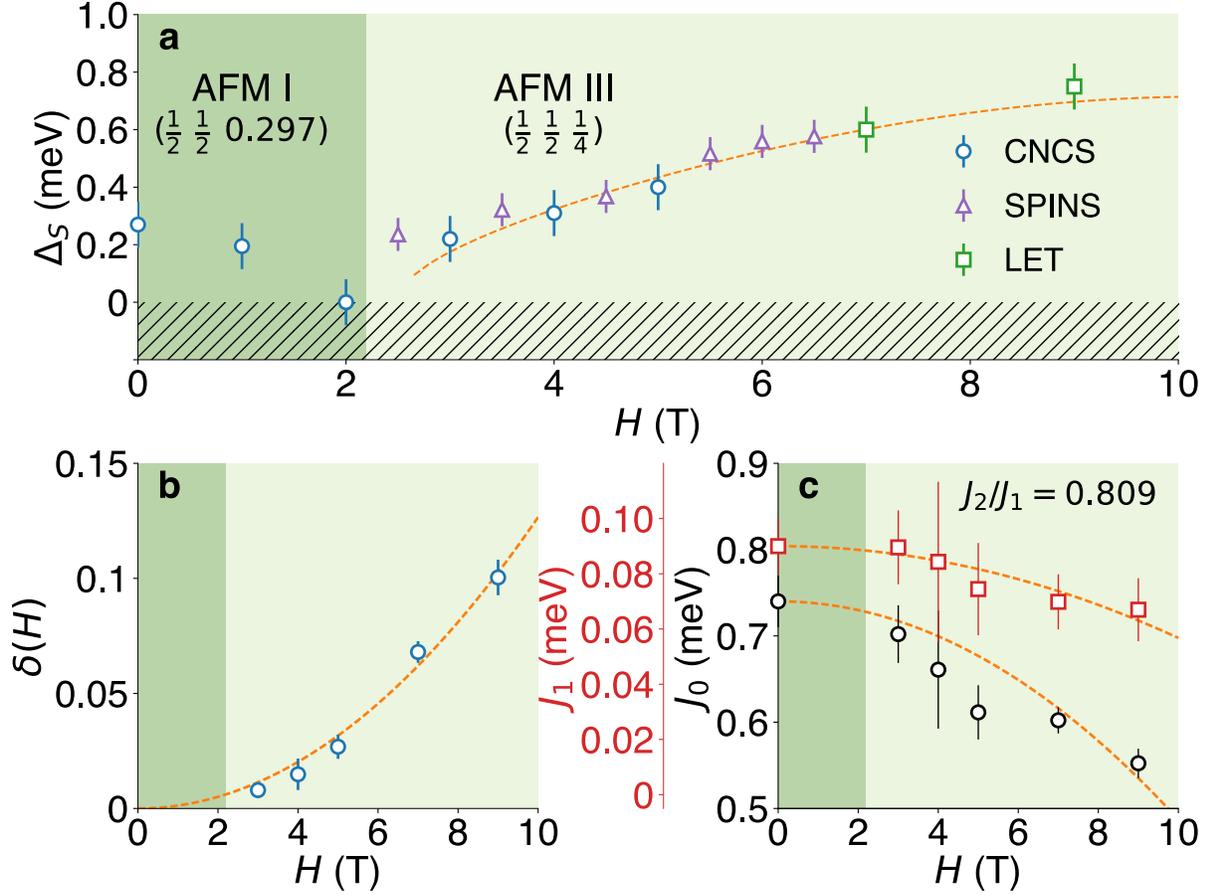

**Figure 4: Salient parameters of the effective spin model related to Axial-Next-Nearest-Neighbor (ANNNI) framework[5] to describe the field-tuned uniaxial anisotropy in CeRhIn$_5$.** **a** Spin gap $\Delta_S$ extracted from spin wave spectra measured via neutron spectroscopy (cf. Fig. 3) as function of magnetic field $H$. Squares indicate the gap was measured on LET at $T = 2$ K, circles on CNCS at $T = 2$ K, and triangles on SPINS at $T = 0.3$ K. Error bars reflect instrument resolution. The dark and light backgrounds denote the boundary between the incommensurate helical AFM order (AFM I, $k_\mathrm{I} = (1/2\ 1/2\ 0.297)$) and the commensurate sine-square wave AFM order (AFM III, $k_\mathrm{III} = (1/2\ 1/2\ 1/4)$), below and above $H_\mathrm{III} = 2.1$ T, respectively. The dashed line is a fit to the gap function $\Delta_S(H)$ derived from the ANNNI model. The finite gap at $H = 0$ is not due to ANNNI physics.[15] **b** $H$-dependence of the magnitude of the uniaxial magnetic anisotropy $\delta$. The dashed line denotes $\delta(H) = I_\delta H^2$ with $I_\delta = 0.0013(1)$ 1/T$^2$. **c** $H$-dependence of the nearest-neighbor magnetic exchange integrals $J_0$ and $J_1$. Next-nearest-neighbor exchange integral $J_2$ scales as $0.809\ J_1$ and is therefore not shown.[15] All error bars in **b** and **c** represent standard deviations.

Supplementary Information for "**Tunable Emergent Heterostructures in a Prototypical Correlated Metal**"

D. M. Fobes,[1] S. Zhang,[2] S.-Z. Lin,[3] Pinaki Das,[1,*] N. J. Ghimire,[1,§] E. D. Bauer,[1] J. D. Thompson,[1] L.W. Harriger,[4] G. Ehlers,[5] A. Podlesnyak,[5] R.I. Bewley,[6] A. Sazonov,[7] V. Hutanu,[7] F. Ronning,[1] C. D. Batista,[1,2] and M. Janoschek[1,†]

[1]MPA-CMMS, Los Alamos National Laboratory, Los Alamos, New Mexico 87545, USA
[2]Department of Physics and Astronomy, The University of Tennessee, Knoxville, Tennessee 37996, USA
[3]T-4, Los Alamos National Laboratory, Los Alamos, New Mexico 87545, USA
[4]NIST Center for Neutron Research, National Institute of Standards and Technology, Gaithersburg, Maryland 20899, USA
[5]QCMD, Oak Ridge National Laboratory, Oak Ridge, Tennessee 37831, USA
[6]ISIS Facility, STFC Rutherford Appleton Laboratory, Harwell Science and Innovation Campus, Chilton, Didcot, Oxon OX11 0QX, United Kingdom
[7]Neutronenforschungsquelle Heinz Maier-Leibnitz FRM II TU Munich, Lichtenbergstr. 1, D-85747 Garching, Germany

[*]Current address: Division of Materials Sciences and Engineering, Ames Laboratory, U.S. DOE, Iowa State University, Ames, Iowa 50011, USA
[§]Current address: Argonne National Laboratory, Lemont, Illinois 60439, USA
[†]Corresponding author: mjanoschek@lanl.gov

**Dependence of the spin wave spectra in CeRhIn$_5$ as function of magnetic field**

In addition to the data presented in Fig. 3 at $H = 7$ T, full spin wave spectra were also obtained for applied magnetic fields of $H = 3, 4, 5,$ and 9 T. Data at all fields, for three slices along high symmetry directions ($h$00), ($hh$0) and (00$l$), are shown in Fig. S1. Two clear trends as a function of increasing field are apparent from these data: the magnitude of the spin wave gap increases, indicating a change in the uniaxial anisotropy as a function of field, and the bandwidth of spin waves decreases, indicating that the exchange integrals lessen with field.

To obtain the exchange integrals $J_0$, $J_1$, and $J_2$, and the magnitude of the additional field-dependent uniaxial anisotropy $\delta(H)$, for each magnetic field we simultaneously fit the three shown data slices for $\hbar\omega \gtrsim 0.15$ (above the incoherent scattering line) to a theoretical dynamic magnetic susceptibility $\chi''(\mathbf{q}, \hbar\omega)$ using a standard least-squares technique, as implemented in NeutronPy. $\chi''(\mathbf{q}, \hbar\omega)$ is derived from the Hamiltonian presented in Eq. (1) of the main text, and given by the following equations:

$$\frac{\chi''(\mathbf{q},\hbar\omega)}{(1-e^{-\hbar\omega/k_B T})} = \frac{d^2\sigma}{d\hbar\omega\, d\Omega} = r_0 \frac{k_f}{k_i} f(\mathbf{q})^2 \sum_{\nu=x,y,z}(1-q_\nu^2)\text{Im}[\chi_{\nu\nu}(\mathbf{q},\hbar\omega)], \quad (S1)$$

where

$$\chi_{xx}(\mathbf{q},\hbar\omega) = \frac{1}{2}\left(\tilde{\chi}_{xx}(\mathbf{q}+\mathbf{Q},\hbar\omega) + \tilde{\chi}_{xx}(\mathbf{q}-\mathbf{Q},\hbar\omega)\right) \quad (S2)$$

$$\chi_{yy}(\mathbf{q}, \hbar\omega) = \frac{1}{2}\left(\tilde{\chi}_{yy}(\mathbf{q} + \mathbf{Q}, \hbar\omega) + \tilde{\chi}_{yy}(\mathbf{q} - \mathbf{Q}, \hbar\omega)\right) \quad (S3)$$

$$\chi_{zz}(\mathbf{q}, \hbar\omega) = \frac{1}{2}\tilde{X}_{zz}(\mathbf{q}, \hbar\omega) \quad (S4)$$

and

$$\tilde{\chi}_{xx}(\mathbf{q}, \hbar\omega) = \frac{S}{4}\left[\frac{(1+\cos\phi_q)(u_{Aq}-v_{Aq})^2 2\omega_{1q}}{\omega_{1q}^2+\omega_n^2} + \frac{(1-\cos\phi_q)(u_{Bq}-v_{Bq})^2 2\omega_{2q}}{\omega_{2q}^2+\omega_n^2}\right] \quad (S5)$$

$$\tilde{\chi}_{zz}(\mathbf{q}, \hbar\omega) = \frac{S}{4}\left[\frac{(1-\sin\phi_q)(u_{Aq}+v_{Aq})^2 2\omega_{1q}}{\omega_{1q}^2+\omega_n^2} + \frac{(1+\sin\phi_q)(u_{Bq}+v_{Bq})^2 2\omega_{2q}}{\omega_{2q}^2+\omega_n^2}\right] \quad (S6)$$

$$\tilde{\chi}_{yy}(\mathbf{q}, \hbar\omega)\Big|_{T=0K} = \frac{1}{2}\left[\sum_{\mathbf{k}sr} \frac{1}{2}\left(\frac{u_{s\mathbf{k}+\mathbf{q}}^2 v_{r\mathbf{k}}^2}{\omega_{\mathbf{k}+\mathbf{q}}^s + \omega_{\mathbf{k}}^r - i\omega_n} + \frac{v_{s\mathbf{k}+\mathbf{q}}^2 u_{r\mathbf{k}}^2}{\omega_{\mathbf{k}+\mathbf{q}}^s + \omega_{\mathbf{k}}^r + i\omega_n}\right) + \right.$$
$$\left.\frac{1}{2}\sum_{\mathbf{k}sr} u_{s\mathbf{k}+\mathbf{q}} v_s u_{r\mathbf{k}} v_{r\mathbf{k}} \frac{\omega_{\mathbf{k}+\mathbf{q}}^s + \omega_{\mathbf{k}}^r}{\left(\omega_{\mathbf{k}+\mathbf{q}}^s + \omega_{\mathbf{k}}^r\right)^2 + \omega_n^2}\right] \quad (S7)$$

in which $\hbar\omega_{(1,2)}(\mathbf{q})$ is given by Eq. (2) in the main text, $i\omega_n = \omega + i\alpha$, and

$$u_{(A,B)\mathbf{q}} = \sqrt{\frac{1}{2}\left(\frac{|J_{0\mathbf{q}}^N + J_{2\mathbf{q}}^N \pm |J_{1\mathbf{q}}||}{\omega_{(1,2)\mathbf{q}}} + 1\right)}, \quad (S8)$$

$$v_{(A,B)\mathbf{q}} = \text{sign}\left(\frac{J_{0\mathbf{q}}^N + J_{2\mathbf{q}}^N \pm |J_{1\mathbf{q}}|}{J_{0\mathbf{q}}^A + J_{2\mathbf{q}}^A \pm |J_{1\mathbf{q}}|}\right)\sqrt{\frac{1}{2}\left(\frac{|J_{0\mathbf{q}}^A + J_{2\mathbf{q}}^A \pm |J_{1\mathbf{q}}||}{\omega_{(1,2)\mathbf{q}}} - 1\right)} \quad (S9)$$

where

$$J_{0\mathbf{q}}^N = J_0 S(\Delta_0 - 1 + \delta_0)\sum_{\nu=x,y}\cos(2\pi q_\nu) + 4J_0 S(1+\delta_0), \quad (S10)$$

$$J_{2\mathbf{q}}^N = J_2 S(\Delta_2 - 1 + \delta_2)\cos(4\pi q_z) + 2J_2 S(1+\delta_2), \quad (S11)$$

$$J_{0\mathbf{q}}^A = J_0 S(\Delta_0 + 1 - \delta_0)\sum_{\nu=x,y}\cos(2\pi q_\nu), \quad (S12)$$

$$J_{2\mathbf{q}}^A = J_2 S(\Delta_2 + 1 - \delta_2)\cos(4\pi q_z), \quad (S13)$$

$$J_{1\mathbf{q}} = J_1 S(\Delta_1 \cos(2\pi q_z) - i(1-\delta_1)\sin(2\pi q_z)) \quad (S14)$$

The derivation of these equations is discussed in detail below; here we will discuss the details of the least squares fitting. The values $\Delta = \Delta_0 = \Delta_1 = \Delta_2$ and $\alpha$ are fixed to quantities obtained previously at $H = 0$ T.[1] $\Delta = 0.82$ is the magnitude of the easy-plane exchange anisotropy. $\alpha = 0.15$ is a phenomenological damping constant. The vector $\mathbf{q} = \mathbf{k}_{III} = (1/2\ 1/2\ 1/4)$ is the magnetic ordering wave vector in the AFM III phase. The ratio $J_2/J_1 = 0.809$, determined theoretically for the AFM I phase, was fixed for fitting of the field-dependent data because (1) in least squares fits large changes to the ratio did not accurately reproduce the spin wave spectra, but small variations to the ratio were outside of the resolution, and (2) the ratio was not theoretically expected to exhibit a significant field-dependence up to 9 T. Furthermore, because the magnetic field was applied along (110), the easy-plane anisotropy $\Delta$ is not expected to vary; to confirm this assumption we performed least-squares fitting with $\Delta$ as a free parameter, and observed no changes within the error bars. Therefore, $J_0$, $J_1$, and $\delta$ were the free parameters in the fit. Because of the small variations between the different $\delta_{0,1,2}$, and inability to uniquely distinguish between them in the fit due to finite data resolution, we assumed $\delta_0 = \delta_1 = \delta_2$. In Fig. S1 we also show the resulting theoretical dynamic magnetic susceptibility $\chi''(\mathbf{q}, \hbar\omega)$ from the fits of each of the data, as described above. Additionally, constant-**q** cuts through the magnetic zone center $\mathbf{k}_{III}$ and the magnetic zone boundary $\mathbf{q} = (1/4,1/4,1/4)$ as a function of energy transfer $\hbar\omega$ for the $H = 7$ T data is shown in Fig. S2. At the zone center two peaks representing distinct spin wave branches are clearly visible. Here the upper peak actually contains two branches that cannot be distinguished within our experimental resolution. The existence of three spin wave branches per each of the two dispersion solutions is a result of the small magnetic Brillouin zone.[1]

To further characterize the evolution of the spin wave gap with magnetic field we performed a series of inelastic neutron measurements using a triple-axis spectrometer. Constant-**q** scans were performed at $\mathbf{k}_{III}$ for $H > 2.1$ T at $T = 0.3$ K as a function of energy transfer with a fixed $E_f$. The spin wave gap was determined by least-squares fits to Gaussian functions, as shown in Fig. S3. Data points resulting from these fits are shown in Fig. 4a. These data points were subsequently fit to the equation

$$\Delta_S = \sqrt{2\delta(2J_0 + J_2)[(2J_0 + J_2)(1 + \delta + \Delta) - J_1\Delta]} \qquad (S15)$$

derived from Eq. (2) in the main text. The exchange integrals $J$ and the uniaxial-anisotropy terms are both magnetic field dependent, with $J = J(H = 0\text{ T}) - I_J H^2$, and $\delta = I_\delta H^2$. Separate fitting to the gap function (cf. Fig. 4a) and the values of $\delta$ derived from fitting the whole spectra (cf. Fig 4b) result in $I_\delta = 0.0014(1)$ and $0.0013(1)$, with $\chi^2_{reduced} = 2.86$ and $1.14$, respectively.

**Temperature dependence of the propagation vector in the elliptical helix phase (AFM II)**
To determine the evolution of the ordering wave vector near the AFM II to AFM III phase boundary we performed a series of high resolution triple-axis diffraction measurements on SPINS. The instrument in triple-axis mode with a flat analyzer, collimations of 20'-S-10' and a cooled Be-filter were used to obtain the best possible **q**(momentum)-resolution. Diffraction scans along the (00$l$) direction were performed at a constant field $H = 3.5$ T as a function of decreasing temperature $T$. The wave vector was determined by least-squares fitting to a Gaussian function with a constant background, as shown in Fig. S4. Data points resulting from these fits are shown in Fig. 2d.

**The ordered magnetic moment and Kondo interaction of CeRhIn$_5$**
To accurately determine the ordered magnetic moment of CeRhIn$_5$, a necessary quantity to generate the theoretical phase diagram as shown in Fig. 2b, we performed neutron diffraction measurements on a sample optimized for hot neutrons ($\lambda = 0.7$ Å). Previous values for the ordered moment vary between $0.26\ \mu_B$ and $0.75\ \mu_B$;[2-7] this inconsistency between measured values is likely attributed to the larger incident wavelengths used, $\lambda > 1.28$ Å, which result in larger neutron absorption by Rh and In, necessitating complicated absorption corrections. To mitigate these issues, we utilized incident wavelength $\lambda = 0.7$ Å, resulting in a 1/$e$ absorption length of ~4 mm, therefore allowing for a larger single crystal sample volume of ~4 mm x 4 mm x 21 mm. A total of 149 structural and 53 magnetic peaks were scanned; The experimental structure factors of the measured Bragg reflections were obtained with DAVINCI [22]. Refinement was performed using single crystal refinement implemented in FullProf,[8] and refining to the established nuclear (P4/*mmm*)[9] and magnetic (helical)[2] structures, we obtained an ordered moment of $m = 0.54(2)\,\mu_B$. Measurements were performed at $T = 1.5$ K, below the saturation temperature of the magnetic order parameter. Complete details may be found in Ref. 10.

This result for the size of the ordered magnetic moment suggests that the Kondo interaction in CeRhIn$_5$ is sizable. Notably, for a total absence of Kondo screening in CeRhIn$_5$, the ordered magnetic moment is expected to be $0.92\,\mu_B$, by calculation from crystalline electric field (CEF) excitations.[11] Although transverse spin fluctuations may also reduce the magnitude of the ordered moment, based on our effective spin Hamiltonian for the ground state of CeRhIn5,[1] we find that spin fluctuations alone would only reduce the ordered moment by 17% compared to the full moment. In contrast, the measured ordered magnetic moment ordered moment $m = 0.54(2)\,\mu_B$ is reduced 41% compared to $0.92\,\mu_B$, indicating significant Kondo screening. This is further supported by the deviation of the measured magnetic form factor from a pure Ce$^{3+}$ magnetic form factor.[10]

**Mean Field result for effective spin Hamiltonian**
At mean-field level, the energy of the single-Q AFM I phase is given by

$$E_{AFM-I} = \left(-2J_0 - J_2 - \frac{J_1^2}{8J_2}\right)NS^2, \qquad \text{(S16)}$$

and the energy of the AFM III phase is given by

$$E_{AFM-III} = (-2J_0 - J_2)(1 + \delta)NS^2. \qquad \text{(S17)}$$

The critical value of the exchange anisotropy $\delta_c$ is obtained from the condition $E_{AFM-I} = E_{AFM-III}$, which leads to $\delta_c^{MF} \sim 0.0091$. In this simplified mean field analysis, we are not including the higher-harmonics which are induced at finite field (finite easy-axis anisotropy). The unconstrained mean field treatment leads to $\delta_c^{CEF} = 0.012$, which is very close to the previous estimate.

**Derivation of the spin wave spectrum**
We first consider the spin wave spectrum of the collinear phase AFM III. We rotate the local reference frame of each spin operator, $S_j \to \tilde{S}_j$, in such a way that the magnetic ordering becomes ferromagnetic in the new reference frame. Due to this transformation, the spin exchange within the tetragonal basal plane ($J_0$), and the next nearest neighbor spin exchange along the *c*-axis ($J_2$) become ferromagnetic, while the sign of the nearest neighbor spin exchange along the *c*-axis ($J_1$) alternates between consecutive bonds. Thus, in the new reference frame the magnetic unit cell consists of two neighboring sites along *c*-axis. We will refer to these two sites as the A and B sublattices. Correspondingly, the magnetic Brillouin zone is twice smaller along the *c*-axis.

The magnon spectrum is obtained by applying linear spin-wave theory, *i.e.* the spin operators $\tilde{S}_j$ are expressed in terms of Holstein-Primakoff (H-P) bosons:

$$\tilde{S}_j^+ = \tilde{S}_j^z + i\tilde{S}_j^x = \sqrt{2S - n_b(j)}\, b_j, \quad (S18)$$

$$\tilde{S}_j^- = \tilde{S}_j^z - i\tilde{S}_j^x = b_j^\dagger \sqrt{2S - n_b(j)}, \quad (S19)$$

$$\tilde{S}_j^y = \tilde{S}_j^y = S - n_b(j), \quad (S20)$$

where $n_b(j) = b_j^\dagger b_j$ and we have assumed that the field induced easy-axis is along the *y*-direction. After substituting this representation into the effective spin Hamiltonian (Eq. (1) in the main text) and keeping terms up to quadratic level in the (H-P) boson operators, we obtain the following spin wave Hamiltonian

$$\mathcal{H}_{sw} = \sum_q \begin{pmatrix} b_{A\mathbf{q}}^\dagger \\ b_{B\mathbf{q}}^\dagger \\ b_{A-\mathbf{q}} \\ b_{B-\mathbf{q}} \end{pmatrix}^T \begin{pmatrix} \frac{J_{0\mathbf{q}}^N + J_{2\mathbf{q}}^N}{2} & \frac{J_{1\mathbf{q}}}{2} & \frac{J_{0\mathbf{q}}^A + J_{2\mathbf{q}}^A}{2} & \frac{J_{1\mathbf{q}}}{2} \\ \frac{J_{1\mathbf{q}}^*}{2} & \frac{J_{0\mathbf{q}}^N + J_{2\mathbf{q}}^N}{2} & \frac{J_{1\mathbf{q}}^*}{2} & \frac{J_{0\mathbf{q}}^A + J_{2\mathbf{q}}^A}{2} \\ \frac{J_{0\mathbf{q}}^A + J_{2\mathbf{q}}^A}{2} & \frac{J_{1\mathbf{q}}}{2} & \frac{J_{0\mathbf{q}}^N + J_{2\mathbf{q}}^N}{2} & \frac{J_{1\mathbf{q}}}{2} \\ \frac{J_{1\mathbf{q}}^*}{2} & \frac{J_{0\mathbf{q}}^A + J_{2\mathbf{q}}^A}{2} & \frac{J_{1\mathbf{q}}^*}{2} & \frac{J_{0\mathbf{q}}^N + J_{2\mathbf{q}}^N}{2} \end{pmatrix} \begin{pmatrix} b_{A\mathbf{q}} \\ b_{B\mathbf{q}} \\ b_{A-\mathbf{q}}^\dagger \\ b_{B-\mathbf{q}}^\dagger \end{pmatrix} \quad (S21)$$

$$- 2J_0 S^2 N(1 + \delta_0) - J_2 S^2 N(1 + \delta_2),$$

where the momentum **q** belongs to the reduced Brillouin zone introduced above. Diagonalization of the above spin wave Hamiltonian through usual Bogoliubov transformation gives rise to Eq. (2) of the main text. The spin wave spectrum of the helical magnetic phase (AFM I) is obtained in a similar way.[1]

**Origin of spin gap at zero magnetic field**
Our recent neutron spectroscopy measurements of the spin wave spectrum of CeRhIn$_5$ at zero magnetic field and ambient pressure observed a small spin gap $\Delta_S \approx 0.25$ meV.[1] However, because the magnetic ground state of CeRhIn$_5$ is an *incommensurate* long-period helix, the spin wave spectrum is expected to exhibit a gapless Goldstone mode in the absence of magnetic anisotropies in the tetragonal plane. Notably, although an in-plane $C_4$ anisotropy is possible for a tetragonal structure, the gapless Goldstone mode is still protected by translational symmetry of the magnetic helix along *c*. In principle, a sufficiently strong $C_4$ magnetic anisotropy (comparable to the exchange interactions along *c*), would distort the magnetic helix and break the associated translational symmetry, resulting in the formation of a spin gap. However, distortion of the helix would also lead to higher harmonic diffraction peaks, which have never been observed. Due to the size of the gap, the associated intensity of the peaks resulting from the distortion should, however, make them straightforward to observe. A previous calculation further demonstrated that the presence of crystallographic defects that break the translation invariance of the helix are also too small to explain the gap.[1]

However, the significant Kondo screening in CeRhIn$_5$ that was identified by means of our neutron diffraction experiments, described above and in Ref. 10, provides an alternative explanation of the spin gap $\Delta_S$ at zero magnetic field, because it suggests the presence of strong longitudinal magnetic fluctuations. In analog to insulating quantum magnets, in which longitudinal fluctuations are induced by spin dimerization,[12,13] this Kondo-induced longitudinal mode is expected to be critically damped and corresponds to a massive (gapped) Higgs mode. The spin waves observed in Ref. [1] as well as the spin waves at non-zero magnetic field (see above) are indeed damped. The energy scale that determines the size of the spin gap associated with this longitudinal mode arises from the competition between Kondo and RKKY interactions, and its calculation requires the solution of the Kondo lattice model in the presence of magnetic order, which goes beyond state-of-the-art solid state theory. In general, we can say that the gap must vanish at the pressure-induced quantum critical point where the magnitude of the moment is completely suppressed by the Kondo effect. Careful measurements of the spin wave gap as function of pressure in CeRhIn$_5$ will be useful to confirm the scenario proposed here.

**Details of the calculation of the magnetic phase diagram**

To obtain the phase diagram shown in Fig. 2b of the main text, we first treat spins in Eq. (1) in the main text as classical spins and then solve it using the mean-field approximation. The mean-field free energy functional is $\mathcal{F} = \mathcal{E} - T\,S_e$, with energy $\mathcal{E}$ and entropy $S_e$:

$$\mathcal{E} = \sum_{ij}\left[J_{ij}\left((1-\delta)m_i^x m_j^x + (1+\delta)m_i^y m_j^y\right) + \Delta J_{ij} m_i^z m_j^z\right] - g_J \mu_B h \sum_j m_j^x, \quad (S22)$$

$$S_e = -\sum_i \left[\beta h_i m_i - \ln\left(\frac{\sinh(\beta h_i)}{\beta h_i}\right)\right], \quad (S23)$$

where $m_i$ is the mean-field value of spin $m_i \equiv \langle S_i \rangle_T$, and $\beta = 1/T$. Here, $\langle \cdots \rangle_T$ represents the thermal average. The molecular field $h_i$ is given by

$$\mathbf{m}_i = \left[\coth(\beta h_i) - \frac{1}{\beta h_i}\right]\frac{\mathbf{h}_i}{h_i}, \quad (S24)$$

We then minimize $\mathcal{F}$ by performing numerical annealing via the Markov chain Monte Carlo method,[14] which reduces the chance of trapping in a metastable state, and subsequently use the relaxation method to find the state with minimal free energy. Because the ordering wave vector is temperature-dependent, we continuously change the system size along the c-direction from 4 to 80, and keep the solution with the lowest free energy.

One remarkable experimental observation is that the Néel temperature $T_N$ increases with applied magnetic field. To reproduce the measured phase diagram, we choose a $I_\delta$ corresponding to $\delta_c^{MF} = 0.0091$. This value agrees with the anisotropy extracted from the spin wave spectrum, validating the Hamiltonian in Eq. (1) of the main text.

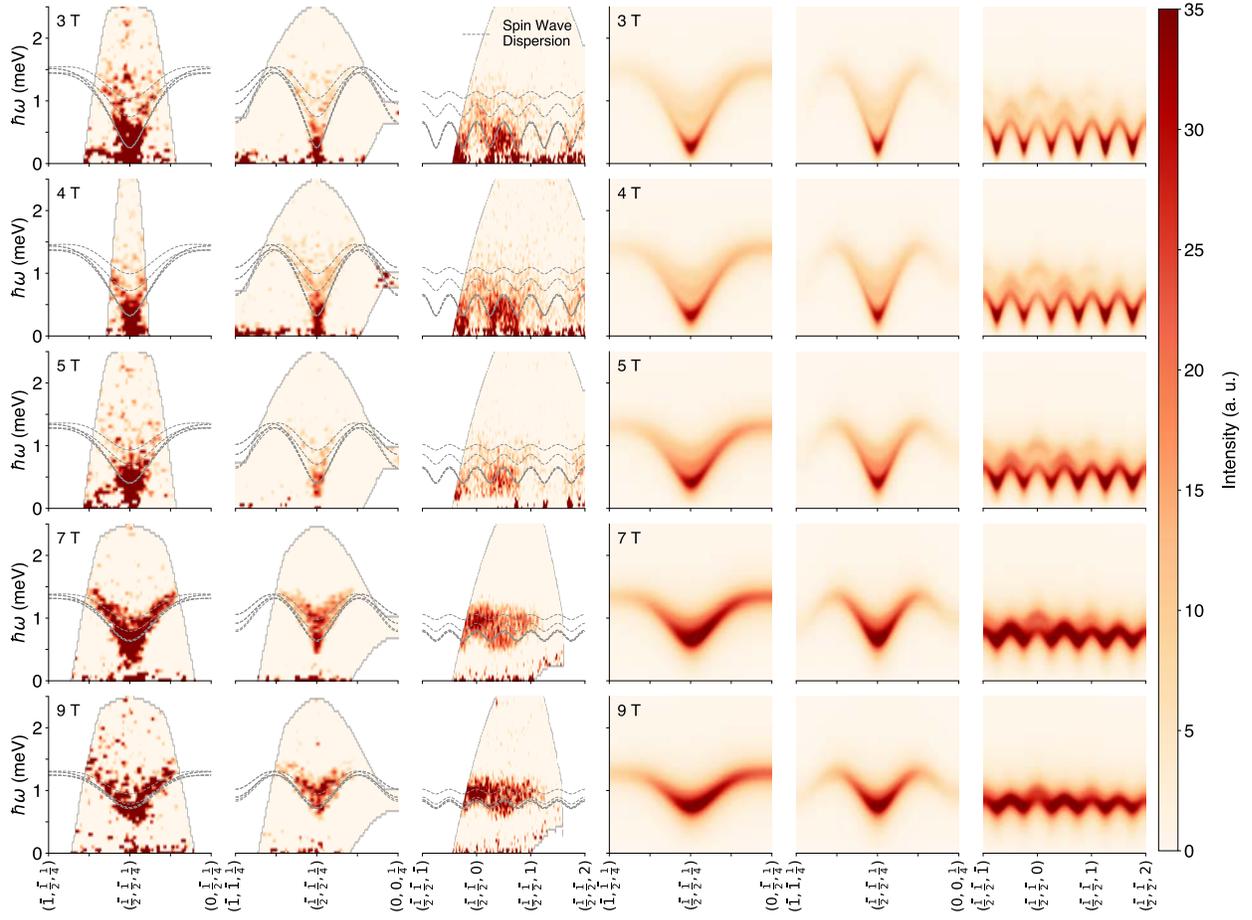

**Figure S1: Experimental and theoretical spin wave spectra in the commensurate sine-square wave antiferromagnetic phase (AFM III).** Measured (3 left columns) and calculated (3 right columns) spin excitation spectra at $H = 3, 4, 5, 7$, and $9$ T in the AFM III phase where $\bm{k}_{\mathrm{III}} = (1/2,1/2,1/4)$, along three high symmetry directions: $(h00), (hh0), (00l)$. Dashed lines indicate the spin wave dispersions from Eq. (2) of the main branch $\omega(\bm{q})$, and two additional branches from Umklapp scattering, $\omega(\bm{q} + \bm{k}_{\mathrm{III}})$ and $\omega(\bm{q} - \bm{k}_{\mathrm{III}})$ resulting from the least-squares fits to the theoretical dynamic susceptibility $\chi''(\bm{q}, \hbar\omega)$.

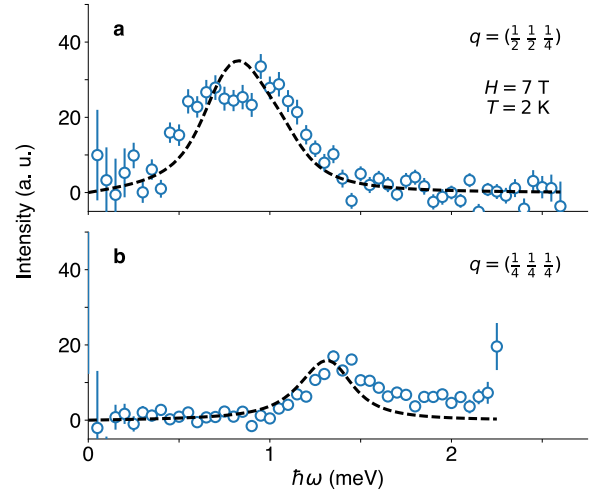

**Figure S2: Constant-q scans through spin wave spectra at 7 T** for **a** $q = k_{III}$ (magnetic zone center) and **b** $q = (1/4\ 1/4\ 1/4)$ (magnetic zone boundary). Data was taken at $T = 2$ K. Dashed black lines are a calculated $\chi''(q, \hbar\omega)$ using the fitted parameters for the 7 T data; a grid of $\chi''(q, \hbar\omega)$ was generated for a range around $q$ reflecting the approximate instrument resolution and collapsed onto $q$. At the zone center two spin wave branches are clearly visible (cf. Fig. S1 and text).

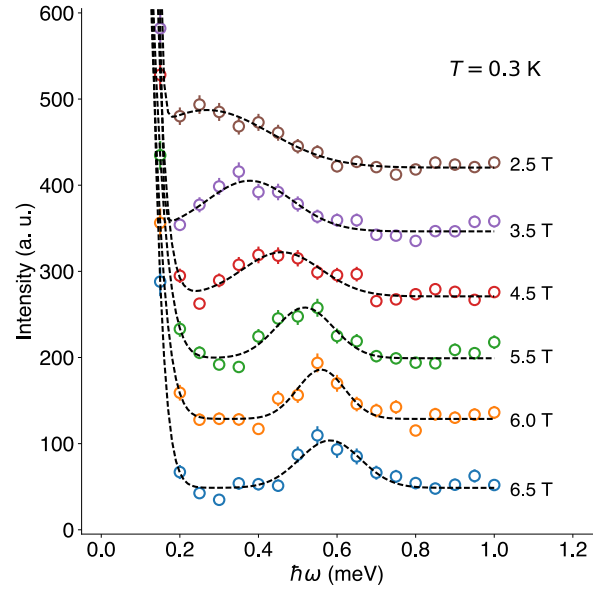

**Figure S3: Evolution of the spin wave gap in the AFM III phase with applied magnetic field.** Energy transfer $\hbar\omega$ scans at $k_{\mathrm{III}}$ at $T = 0.3$ K taken at $H = 2.5, 3.5, 4.5, 5.5, 6$, and $6.5$ T. Dashed lines are fits to two Gaussian functions; the incoherent line is a Gaussian at fixed position 0 meV and full width at half maximum (FWHM) 0.115 meV, reflecting instrument resolution. Data are shifted for clarity.

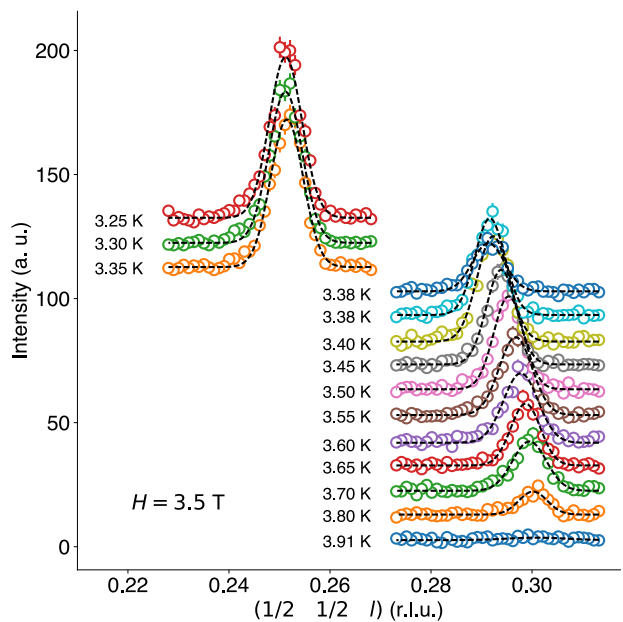

**Figure S4: Magnetic ordering wave vector evolution across the AFM II – AFM III phase transition.** Diffraction scans around $\boldsymbol{k}_{III}$ or $\boldsymbol{k}_I$ along the $(00l)$-direction at $H = 3.5$ T taken for $3.25\,K < T < 3.91$ K. Dashed black lines are single Gaussians with a constant background. Data are shifted for clarity.